# Wavelength Assignment in Design DWDM Transport Network Using Algorithm BCO-RWA

Dmitry Ageyev, Alexander Pereverzev

*Abstract* - **The main problem in designing DWDM transport networks is the wavelength assignment of light paths. One way of solving this problem is to use the algorithm BCO-RWA. However, BCO-RWA has the following disadvantages: algorithm not solved the problem of choosing the location of the optical convector in the network; base algorithm ignores placement optical convector during calculating route selecting probability; base algorithm do not take into account the nonlinear four-wave mixing phenomenon. In this paper we present an algorithm which takes into account a number of disadvantages due to modifications introduced in the algorithm.**

## I. INTRODUCTION

Modern multi-service networks that provide many services require the guaranteeing of high speeds, these requirements correspond to the transport network technology based DWDM. DWDM technology provides the most extensive and cost-effective way to expand the bandwidth of fiber-optic channels to hundreds. Bandwidth of optical links based on DWDM systems can be increased, adding to the extent of network development in existing equipment new optical channels.

Among the tasks that need to be solved in the planning of transport networks based on technologies DWDM, one of the main problems is that the wavelength assignment of light paths. This problem is known as the routing and wavelength assignment (RWA).

The RWA problem takes two forms: static and dynamic. In the static case, all connection requested at network are known in advance, thus a routing decision can be made based on the full information of the traffic to be served by the network. In the dynamic case, a connection request has to be routed and wavelengths assigned independently of other connections, which either have already been assigned.

In this paper concentrate on the static version of RWA problem in which the objective is to maximize the number of established ligthpaths.

To solve this problem in this paper an algorithm was used Bee Colony Optimization Routing and Wavelength Assignment (BCO-RWA) [1]. BCO-RWA – heuristic algorithm based on the basic principles of modeling the behavior of bees to solve optimization problems. The main advantage of this algorithm is - a low computational complexity.

## II. STATEMENT OF THE PROBLEM

The solution of the assignment of wavelengths of light used routes in the network based on WDM technology with BCO-RWA algorithm was considered in [2].

We assume that a given WDM network is a single fiber that means each physical link has one separate fiber for each direction of transmission of traffic. Each fiber supports the same number of available wavelengths W (for DWDM 40).

An optical network composed of N nodes and L optical links can be represented by a corresponding graph with N nodes and L undirected edges. In the network not be two light paths traversing through the same link(i j) will have the identical wavelength assignment to them.

However, the analysis of this work has identified the following imperfections: not solved the problem of choosing the location of the optical convector in the network, when calculating the probability of selecting a route ignore placement optical convector, en route, do not take into account the nonlinear four-wave mixing phenomenon.

To solve the problem of choosing the location of the optical convector modification was introduced in the BCO-RWA algorithm, which is based on finding the bottleneck in the network (node through which the greatest number of established connections) which placement the optical convector.

When calculating the probability of selecting the route algorithm was added to the calculation takes into account the condition, that on the node that is the route may be an optical convector.

## III. THE BCO-RWA ALGORITHM

Artificial bees represent agents, which collaboratively solve complex combinational optimization problems. Every artificial bee generates one solution to the problem. The algorithm consists of two alternating phases: forward pass and backward pass.

At the beginning of the search process all artificial agents are located in the hive. Bees depart from the hive and fly through the artificial network from the left to the right.

Bee's trip is divided into stages. Bee chooses to visit one artificial node at every stage. Each stage represents the collection of all considered origin-destination pairs. Each artificial node is comprised of an origin and destination linked by a number of routes.

Lightpath is a route chosen by bee agent. Bee agent's entire flight is collection of established lightpaths.

During forward pass every bee visits n stages (bee tries to establish n new light paths). In every stage a bee chooses

Dmitry Ageyev - Kharkiv National University of Radioelectroniks, Lenina Av., 14, Kharkov, 61166, UKRAINE, E-mail:dm@ageyev.in.ua
Alexander Pereverzev - Kharkiv National University of Radioelectroniks, Lenina Av., 14, Kharkov, 61166, UKRAINE, E-mail:pereverzev_aa@mail.ru

one of the previously not visited artificial nodes generated by the bee represents one partial solution of the problem considered.

Bee is not always successful in establishing lightpath when visiting artificial node. Bee's success depends on the wavelengths availability on the specific links. In this way, generated partial solutions differ among themselves according to the total number of established lightpaths.

After forward pass, bees perform backward pass, i.e. they return to the hive. The number of nodes n to be visited during one forward pass is prescribed by the analyst at the beginning of the search progress, such that n<<m, where m is a total number of requested lightpaths.

## IV. ANALYSIS OF RESULTS

The results of comparison of old and new methods for solving the assignment of wavelengths are presented in Table 1. Notation in Table 1: nodes – number of nodes in network topology; mean – the average value of the objective function (number of established connection); variance – deviation value of the objective function from mean;time – the average time for solving problem.

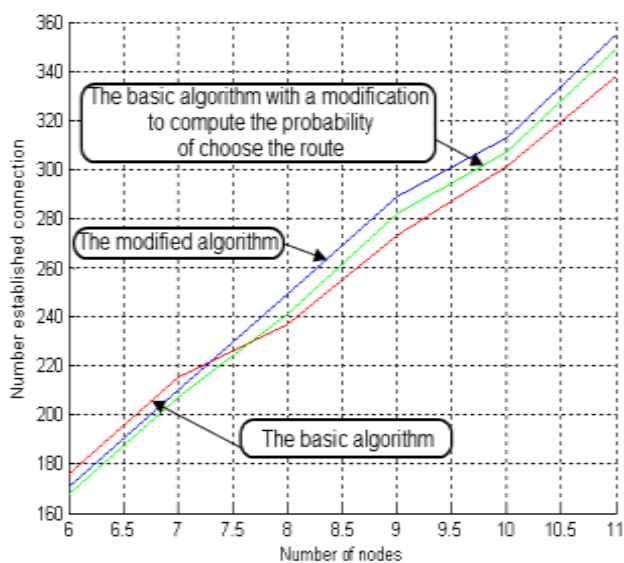

Fig.1 Results of the reseach

A Fig. 1 show that the small dimensions of topologies of 6.7 nodes basic algorithm is better than the modified algorithms.

However, with increasing dimension topologies modified algorithm allows us to establish more connections through the optimal location of the convector than the basic algorithm.

Take into account of four-wave mixing phenomena in the design of DWDM transport network is considered in [3].

Further direction of research is the selection of a quantitative criterion for discarding solutions in four-wave mixing and research the method of using BCO-RWA algorithm in a more than one convector.

TABLE 1
RESULTS OF THE RESEACH

| Nodes | Mean | Variance | Time, s |
|---|---|---|---|
| The basic algorithm | | | |
| 6 | 176 | 33 | 1 |
| 7 | 215 | 79 | 2 |
| 8 | 237 | 241 | 3 |
| 9 | 273 | 113 | 4 |
| 10 | 301 | 66 | 5 |
| 11 | 338 | 457 | 7 |
| The basic algorithm with a modification to compute the probability of choose the route | | | |
| 6 | 168 | 65 | 2 |
| 7 | 207 | 119 | 3 |
| 8 | 241 | 108 | 4 |
| 9 | 282 | 143 | 5 |
| 10 | 307 | 102 | 6 |
| 11 | 349 | 187 | 8 |
| The modified algorithm | | | |
| 6 | 171 | 56 | 2 |
| 7 | 210 | 35 | 3 |
| 8 | 249 | 204 | 4 |
| 9 | 289 | 116 | 5 |
| 10 | 313 | 55 | 6 |
| 11 | 355 | 213 | 8 |

## V. CONCLUSION

As a result of the work of the study we can conclude, that the previously known method of using BCO-RWA algorithm has the following disadvantages: not solved the problem of choosing the location of the optical convector in the network, when calculating the probability of selecting a route ignore placement optical convector, en route, do not take into account the nonlinear four-wave mixing phenomenon.

As a result, modifications of BCO-RWA algorithm was removed a number of deficiencies, and increased efficiency of the algorithm.